\documentclass[aps,pra,onecolumn,superscriptaddress, floatfix]{revtex4-1}
\usepackage{amsmath}
\usepackage{amssymb}
\usepackage{graphicx}
\usepackage{nicefrac}
\usepackage{siunitx}
\usepackage{lipsum}
\usepackage{setspace}
\usepackage{wasysym}

\newcommand{\ThreeJ}[6]{\left( \begin{matrix}  #1 & #2 & #3 \\ #4 & #5 & #6 \end{matrix} \right)}

\newcommand{\SixJ}[6]{\left\{ \begin{matrix} #1 & #2 & #3 \\ #4 & #5 & #6 \end{matrix} \right\}}

\newcommand{\braket}[3]{\left< #1 \right| #2 \left| #3 \right>}

\begin{document}

%\preprint{}

\title{Metrology of Rydberg states of the hydrogen atom}


\author{Simon Scheidegger}
\affiliation{Departement of Chemistry and Applied Biosciences, ETH Zurich, Zurich, Switzerland}
\affiliation{Quantum Center, ETH Zurich, Zurich, Switzerland}
\author{Josef A. Agner}
\affiliation{Departement of Chemistry and Applied Biosciences, ETH Zurich, Zurich, Switzerland}
\author{Hansj\"urg Schmutz}
\affiliation{Departement of Chemistry and Applied Biosciences, ETH Zurich, Zurich, Switzerland}
\author{Fr\'ed\'eric Merkt}

\email[]{merkt@phys.chem.ethz.ch}
\affiliation{Departement of Chemistry and Applied Biosciences, ETH Zurich, Zurich, Switzerland}
\affiliation{Departement of Physics, ETH Zurich, Zurich, Switzerland}
\affiliation{Quantum Center, ETH Zurich, Zurich, Switzerland}

\date{\today}

\begin{abstract}
We present a method to precisly measure the frequencies of transitions to high-$n$ Rydberg states of the hydrogen atom which are not subject to uncontrolled systematic shifts caused by stray electric fields. The method consists in recording Stark spectra of the field-insensitive $k=0$ Stark states and the field-sensitive $k=\pm2$ Stark states, which are used to calibrate the electric field strength. We illustrate this method with measurements of transitions from the 2s$(f=0\text{ and } 1)$ hyperfine levels in the presence of intentionally applied electric fields with strengths in the range between 0.4 and \SI{1.6}{\volt\per\centi\meter}. The slightly field-dependent $k=0$ level energies are corrected with a precisely calculated shift to obtain the corresponding Bohr energies $\left(-cR_{\mathrm{H}}/n^2\right)$. The energy difference between $n=20$ and $n=24$ obtained with our method agrees with Bohr's formula within the 10\,kHz experimental uncertainty. We also determined the hyperfine splitting of the 2s state by taking the difference between transition frequencies from the 2s$(f=0  \text{ and }1)$ levels to the $n=20,k=0$ Stark states. Our results demonstrate the possibility of carrying out precision measurements in high-$n$ hydrogenic quantum states. 
\end{abstract}

%\keywords{}

\maketitle

\section{Introduction}
The hydrogen atom is a fundamental two-body quantum system. Studies of its spectrum by experiment and theory have played a key role in the development of the quantum theory \cite{angstrom68a,balmer85a,bohr13a,schroedinger26a,dirac26b} and of quantum electrodynamics \cite{lamb47a,lamb50a,bethe47a,bethe57a}. Spectroscopic measurements of energy intervals between the quantum states of the hydrogen atom have reached exceptional precision and the results can be exactly explained by first-principles calculations and accurately known physical constants such as the Rydberg constant $R_\infty$, the fine-structure constant $\alpha$ and the proton charge radius ${r_{\rm p}}$. The theoretical treatment of the H atom by relativistic quantum mechanics and quantum electrodynamics is indeed so accurate that the comparison with the results of precision measurements in the H atom can serve to determine the values of these constants \cite{tiesinga21a}.\par
In the past years, a significant revision of the values of $R_\infty$ and $r_{\rm p}$ became necessary after a new measurement of the Lamb shift in muonic hydrogen \cite{pohl10a, antognini13a, lensky22a} challenged earlier results from H-atom spectroscopy, a challenge that was referred to as the proton-radius puzzle. This challenge essentially results from the correlation between the $R_\infty$ and $r_{\rm p}$ values which necessitates the combination of at least two transition frequencies in the H atom to determine these constants. The latest CODATA values of $R_\infty$ and $r_{\rm p}$ are based on a combination of multiple results, in which the 1s--2s interval in H \cite{parthey11a,matveev13a} and the Lamb shift in muonic hydrogen \cite{pohl10a} play a central role. Several recent precision measurements in H confirmed the revised values \cite{beyer17b,bezginov19a} whereas others cover the range between the old and the new values of $R_\infty$ and $r_{\rm p}$ \cite{fleurbaey18a,grinin20a,brandt22a}. Measurement of quantities that are only sensitive to either $r_{\rm p}$, such as electron-scattering measurements \cite{belushkin07a,bernauer10a,sick12a,lee15a,higinbotham16a,horbatsch17a,alarcon19a,xiong19a,lin21a}, or $R_\infty$, such as measurements in non-penetrating Rydberg series of H, have regained interest.
\par
Early, remarkable experiments designed to determine $R_\infty$ from transition frequencies between circular states, \textit{i.e.}, states with orbital-angular-momentum quantum number $\ell=n-1$ and magnetic quantum number $m_\ell=\pm\ell$, of high principal quantum numbers in the H atom were carried out in the group of D. Kleppner at MIT \cite{paine92a,lutwak97a,devries02a}, giving values of $R_\infty$ compatible with the recommended CODATA values available at the time \cite{mohr00a}. In that work, the frequencies of $\Delta n = 1$ transitions between circular states of H were measured with 2-3 Hz accuracy at $n$ values around 30. These transition frequencies scale as $2R_\infty/n^3$ and are completely insensitive to the proton size because the Rydberg electron does not penetrate in the core region. The $2/n^3$ sensitivity factor to $R_\infty$ of these measurement is only $\approx 1\times 10^{-4}$ for the transition between the $n=27,\ \ell=26, m_\ell=26$ and $n=28,\ \ell=27, m_\ell=27$, but this disadvantage could be compensated by the fact that circular states are not sensitive to stray electric fields to first order, and through the exceptional control of all aspects of the millimeter-wave-spectroscopic experiments by the MIT team. An $R_\infty$ value with an absolute uncertainty of 69~kHz and a relative uncertainty of $2.1\times 10^{-11}$ was determined \cite{devries02a}, close to the $R_\infty$ uncertainty value of $7.6\times 10^{-12}$ of the 1998 CODATA adjustment. Since this pioneering work, circular Rydberg states of Rb have been proposed as an alternative system to determine $R_\infty$ \cite{ramos17a}. The properties of circular Rydberg states of any atom or molecule are indeed ideally suited to metrology, as illustrated by the use of such states as ultrasensitive electric-field sensors \cite{facon16a}.
\par
If circular Rydberg states are excepted, high Rydberg states are usually not considered to be suitable for precision measurements because of their high sensitivity to stray electric fields (see discussion in, e.g., Refs. \cite{garreau90b, beyer13a}). In the context of metrology in the H atom, this sensitivity has implied that almost all precision experiments involving Rydberg states of H with $n\geq 3$ have targeted states with $n$ values below 12 \cite{debeauvoir97a,beyer17b,fleurbaey18a} and that the measurements required a careful evaluation of the Stark effect on the level structure induced by stray electric fields.
\par
We introduce here an alternative method to determine $R_\infty$ which relies on measuring the spectra of $|m_\ell| =1$ Rydberg states of the H atom in the presence of intentionally applied electric fields. 
Stark states of the H atom exhibit shifts of $\approx 1.5a_0ekn\mathcal{F}$ that are linear in the field strength $\mathcal{F}$ at low fields and proportional to the integer difference $k= n_1 - n_2$ between the quantum numbers $n_1$ and $n_2$ that arise in the solution of the Schr{\"o}dinger equation in parabolic coordinates ($k = 0, \pm1,  \pm 2,\ldots, \pm(n-1-|m_\ell|)$, where $m_\ell$ is the magnetic quantum number associated with the electron orbital motion) \cite{bethe57a,gallagher88a}. 
Consequently, even-$n$, $k=0, |m_\ell|=1$ states are to first order field insensitive, as circular Rydberg states. Their magnetic moments are, however, much smaller than for circular states, which makes them less sensitive to Zeeman shifts by magnetic stray fields. $|m_\ell| = 1$ Stark states do not possess any $s$ character and their $\ell$-mixed wavefunctions are dominated by nonpenetrating high-$\ell$ components; consequently, their spectral positions are also insensitive to the proton size. Experimentally, we measure the frequencies of transitions from the 2s($f=0$ and 1) states to $n=20,k=0,\pm 2, |m_\ell|$ Stark states and use the separation between the $k=\pm 2$ states to precisely determine the value of the applied field. We then extract the position of the $k=0$ state to determine the Bohr energy $\left(-hcR_{\rm H}n^{-2}\right)$ after correcting for the quadratic Stark shifts. To obtain a value of $R_\infty$ without having to consider its correlation with $r_{\rm p}$, the position of the 2s levels and the  $n=20,k=0,\pm 2, |m_\ell|$ Stark states can be related to the position of the 2p levels using the 2s Lamb shift determined by Bezginov {\it et al.} \cite{bezginov19a}. The sensitivity factor of the measurement to $R_\infty$ is thus 1/4, \textit{i.e.}, more than 3000 times higher than for the measurement based on circular states at $n\approx 30$. Consequently, an accuracy of about 20~kHz would make this measurement competitive with the MIT measurements and we believe that this is achievable. The price to pay for this advantage is that the transition frequencies are in the UV range of the electromagnetic spectrum rather than in the millimeter-wave range and, therefore, compensation of the Doppler effect becomes much more critical. 
\par
In this article, we present several of the key aspects of this method of determining $R_\infty$. We are still in the middle of the data-acquisition process, and use subsets of the data to discuss systematic uncertainties in the measurements of $nkm_\ell\leftarrow 2{\rm s}$ transition frequencies originating from the Stark effect. We also present the determination of (i) the $f=0-f=1$ hyperfine interval in the 2s state, which we obtain by combining two sets of measurements, from 2s($f=0$) and 2s($f=1$) to $n=20$ Stark states, and (ii) the difference between the $n=20$ and $n=24$ Bohr energies by combining measurements from the 2s$(f=1)$ hyperfine state to $n=20$ and 24 Stark states. The article is structured as follows: Section~\ref{sec:experimental_setup} describes the experimental setup and provides details on the laser systems used to prepare H atoms selectively in the 2s($f=0$ and 1) hyperfine states and to record spectra of the $nkm_\ell\leftarrow 2{\rm s}(f)$ transitions, as well as the detection system and the procedure we follow to cancel the Doppler shifts. Section~\ref{sec:Stark_theory} describes how we calculate the energies of the Stark states of H and draws attention to the aspects that are most relevant for the determination of the Bohr energies. Section~\ref{sec:Result} illustrates the current status of our measurements by using small data sets to compare spectra recorded at different electric fields from the two hyperfine components of the 2s state and to $n=20$ and 24 Stark states. The results we present here only concern small energy intervals $(\sim\SI{177}{\mega\hertz}$ for the 2s($f=1\leftarrow f=0$) interval and \SI{2.51}{\tera\hertz} for the difference between the Bohr energies at $n=20$ and 24) obtained by building differences of (currently still blinded) UV laser frequencies. Absolute transition frequencies will be reported when the analysis of the systematic errors related to the Doppler effect is completed. In the last section, we draw several conclusions concerning our new approach.\\
\section{Experimental setup}\label{sec:experimental_setup}
\begin{figure*}[h!]
		\includegraphics[scale=1]{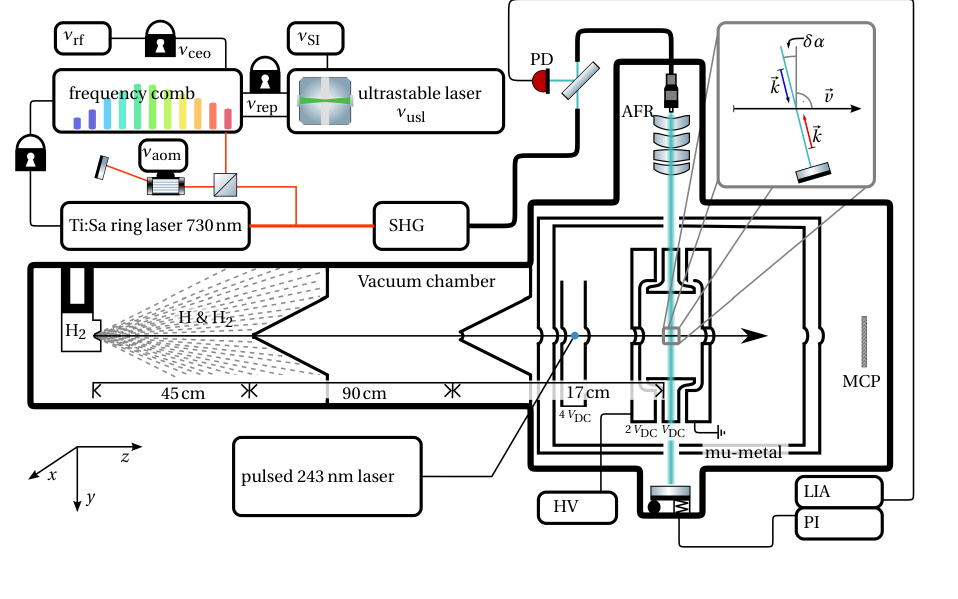}
		\caption{Schematic representation of the experimental setup. Upper part: laser system and geometry of the photoexcitation from the metastable 2s state of H to Rydberg states. Lower part: vacuum chambers in which the supersonic beam of H atoms is generated, these atoms are photoexcited to Rydberg states and the Rydberg states are detected by pulsed field ionization Top right inset: Configuration of laser and supersonic beams used for the determination of Doppler-free frequencies. See text for details.}
		\label{fig:Experimental_setup}
	\end{figure*}

The experimental setup is presented schematically in Fig. \ref{fig:Experimental_setup}. It consists of (i) a differentially pumped set of vacuum chambers in which the H atoms are produced and entrained in a pulsed supersonic beam and subsequently photoexcited to Rydberg states via the metastable 2s state within a double-layer mu-metal magnetic shield; (ii) a pulsed near-Fourier-transform-limited laser system delivering radiation at \SI{243}{\nano\meter} to drive the $2{\rm s}\leftarrow1{\rm s}$ transition; and (iii) an SI-traceable single-mode continuous-wave (cw) UV laser to further excite the H atoms to Rydberg states. The experiment is run in a pulsed mode at a repetition rate of \SI{25}{\hertz}.\par
The hydrogen atom source has been described in Ref. \cite{scheidegger22a} to which we refer for details.\
The hydrogen atoms are produced by dissociating molecular hydrogen in a dielectric-barrier discharge near the orifice of a pulsed cryogenic valve and are entrained in a supersonic beam of H$_2$.  The temperature ($T_0$) of the valve can be adjusted between \SI{45}{\kelvin} and \SI{160}{\kelvin} to vary the forward velocity of the supersonic expansion between \SI{970}{\meter\per\second} and \SI{1800}{\meter\per\second}. The final longitudinal temperature of the supersonic beam ($\approx$\SI{12}{\milli\kelvin}) and its forward velocity ($v_{\text{z}} \approx \sqrt{\nicefrac{2k_{\text{B}}T_{0}\gamma}{m_{\text{H}_2} (\gamma -1)}}$) can be well approximated using the model of an adiabatic expansion \cite{scoles88a}. At valve temperatures below the characteristic rotational temperature  of the carrier gas H$_2\, (\theta_{\text{rot}} \approx$ \SI{90}{\kelvin}), the heat capacity ratio $\gamma$ can be approximated by the one of a monoatomic gas, \textit{i.e.}, $\gamma=\nicefrac{5}{3}$.  The central part of the supersonic beam is selected by two skimmers with diameters of \SI{2}{\milli\meter} and \SI{3}{\milli\meter}  placed at distances of  \SI{45}{\centi\meter} and \SI{135}{\centi\meter} from the nozzle orifice, respectively.\par 
The skimmed supersonic beam enters a magnetically shielded chamber in which the H atoms are excited to Rydberg states in a sequential three-photon absorption process. The $2\text{s}\leftarrow 1\text{s}$ transition is first induced between two copper plates kept at the same electric potential of 4V$_{\text{DC}}$ by the third-harmonic ($\lambda = \SI{243}{\nano\meter}$) beam of a pulse-amplified near-Fourier-transform-limited Ti:Sa laser\,\cite{seiler05a} which crosses the supersonic beam at right angles. The molecular beam then traverses a region with a weak homogeneous electric field $\mathcal{F}_{\text{DC}}=\nicefrac{V_{\text{DC}}}{\text{cm}^{-1}}$, where it intersects a single-mode cw UV laser ($\lambda\approx \SI{368}{\nano\meter}$) used to excite the metastable H(2s) atoms to specific Rydberg-Stark states. These states are field ionized by a large pulsed electric field (up to \SI{6}{\kilo\volt\per\centi\meter}) and the resulting protons are accelerated towards a micro-channel-plate (MCP) detector. The different components are discussed in more details in the following subsections. Spectra of Rydberg-Stark states are recorded by monitoring the H$^+$ field-ionization yield as a function of the UV laser frequency. 

\subsection{Laser system for the 2s $\leftarrow$ 1s transition}\label{sec:2s_excitation}
The 243-nm radiation used to excite the H atoms to the 2s state by nonresonant two-photon excitation is generated by amplification of the 120-ns-long chopped output of a titanium-sapphire (Ti:Sa) seed laser at \SI{729}{\nano\meter} using a Nd:YAG-pumped Ti:Sa multipass amplifier, as described in Ref. \cite{seiler05a}. The output pulses, with pulse energies of $\approx\SI{15}{\milli\joule}$, are frequency tripled in two successive $\beta$-barium-borate (BBO) crystals, resulting in 40-ns-long pulses at \SI{243}{\nano\meter} with typical pulse energies of \SI{800}{\micro\joule}. The 243-nm-laser beam is focused slightly beyond the supersonic beam using a 30-cm-focal-length lens. The use of two skimmers reduces the Doppler width of the $2{\rm s}\leftarrow 1{\rm s}$ transition and enables the full resolution of the $f=0\leftarrow f=0$ and $f=1\leftarrow f=1$ hyperfine components.\par 
Because the 243-nm laser beam propagates along the \textit{x} axis (see Fig. \ref{fig:Experimental_setup}), perpendicularly to both the supersonic beam and the cw UV laser, the focus selects a narrow cylinder (diameter of \SI{0.1}{\milli\meter}) of H atoms with a reduced velocity distribution along the \textit{y} axis (see axis system in Fig. \ref{fig:Experimental_setup}). This selection narrows down the Doppler width of the Rydberg-excitation spectra from the 2s level. The photoexcitation only excites H(1s) atoms in a very restricted longitudinal phase-space volume. Consequently, the H(2s)-atom cloud remains compact and hardly expands as the beam propagates through the 4-cm-long distance separating the $2{\rm s}\leftarrow 1{\rm s}$ excitation region from the $nkm\leftarrow 2{\rm s}$ excitation region. However, the spatial and velocity selection can lead to a nonthermal velocity distribution, potentially resulting in asymmetric Doppler profiles in the Rydberg-excitation spectra.\par
The 243-nm laser unavoidably ionizes a significant fraction of the H(2s) atoms \cite{haas06a}. To avoid stray fields from the generated protons, they are accelerated out of the H(2s) cloud by the electric field $\mathcal{F}_{\text{DC}}$ resulting from the potentials applied between the different electrodes within the mu-metal magnetic shield (see Fig. \ref{fig:Experimental_setup}). To eliminate line broadening caused by interactions between closely spaced Rydberg atoms in the sample volume, the measurements are carried out in a regime where at most one Rydberg atom is in the excitation volume and on average much less than one  field-ionization event is detected per experimental cycle. \par
\subsection{Laser system for the \textit{nkm}$\leftarrow$2s excitation}\label{sec:cw-laser}
The primary laser used for the precision spectroscopy of the $nkm \leftarrow 2{\rm s}$ transition is a commercial continuous-wave (cw)  Ti:Sa ring laser (Coherent, 899-21) pumped by a \SI{12}{\watt} solid-state laser (Coherent, Verdi V-12). The Ti:Sa ring laser is operated in the range  $729\--$\SI{736}{\nano\meter} and provides \SI{1}{\watt} of output power. In addition to the standard actuators of the ring laser, an intra-cavity electro-optic modulator (EOM) (QUBIG, PS3D-BC) is used as a fast actuator to maintain a phase lock to an ultrastable reference laser, as discussed below. Around \SI{98}{\percent} of the optical power is sent to a home-built second-harmonic-generation enhancement cavity (SHG)  equipped with a 12-mm-long lithium-triborate (LBO) crystal cut at Brewster's angle. The SHG cavity is stabilized using a H\"ansch-Couillaud scheme \cite{hansch80a}. The typical conversion efficiency to the second harmonic is \SI{20}{\percent}. The 368-nm output of the SHG cavity is coupled into an optical fiber and guided  to an actively stabilized retroreflector (AFR) setup for Doppler-shift compensation (see below). The forward-propagating and retroreflected laser beams cross the molecular beam at right angles \SI{4}{\centi\meter} downstream of the $2{\rm s}\leftarrow 1{\rm s}$ excitation spot.\par
\begin{figure}[h!]
	\includegraphics[scale=0.9]{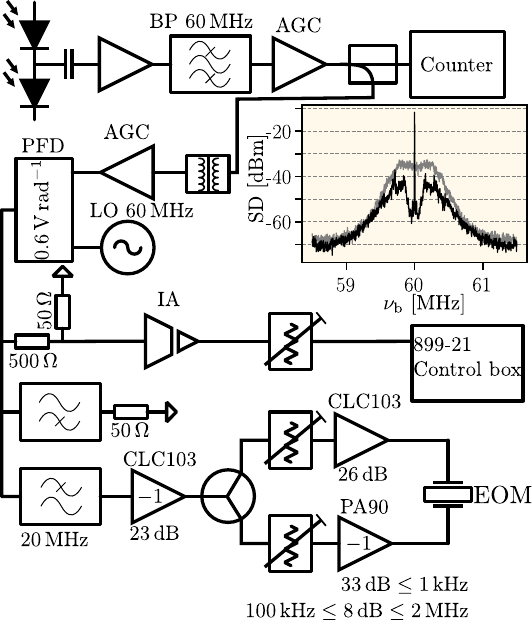}
	\caption{Schematic electric-circuit diagram of the laser-stabilization electronics (see text for details). Color-shaded inset: Spectral density (SD) of the in-loop beat note $\nu_{b}$ recorded with a bandwidth of \SI{3}{\kilo\hertz} with (black) and without (gray) active stabilization using the intracavity EOM.}
	\label{fig:Ti:Sa_lock}
\end{figure}

 
The remaining \SI{2}{\percent} of the fundamental laser power is used for the frequency calibration and stabilization. The light is tightly collimated and sent through an acousto-optic modulator (AOM) (Isomet, M1260-T350L). The first-order diffraction is retro-reflected and its polarization turned by \SI{90}{\degree}, as illustrated in the upper left part of Fig.~\ref{fig:Experimental_setup}. The double-pass configuration induces a shift of the fundamental frequency $\nu_{\text{L}}$ by $2\nu_{\text{aom}}$ which can be adjusted up to \SI{320}{\mega\hertz}. A polarizing beam splitter then deflects the frequency-shifted radiation and sends it through an optical fiber to an amplified, spectrally broadened, and frequency-doubled optically stabilized ultra-low-noise frequency comb (MenloSystems, FC1500-ULN \& M-VIS). The repetition rate of the frequency comb is locked to an ultrastable laser, the frequency of which is referenced to an SI-traceable frequency standard, as characterized in Ref. \cite{husmann21a}. The output of the spectrally broadened frequency comb is dispersed with a reflective grating and the spectral components around $\nu_{\text{L}}$ are selected and spatially overlapped with the laser. The beat, with frequency
\begin{equation}
	\nu_{b} = \nu_{c} - \nu_{\text{L}^\prime}
\end{equation}
between the shifted laser frequency $\nu_{\text{L}^\prime} = \nu_{\text{L}} +2\nu_{\text{aom}}$ and the spectrally closest frequency-comb tooth $\nu_{c}$ is recorded using a balanced photodiode (Thorlabs, PDB425A-AC) and processed using the electronic circuit depicted in Fig. \ref{fig:Ti:Sa_lock}. A bandpass filter centered at \SI{60}{\mega\hertz} is used to suppress beat frequencies originating from neighboring comb teeth. The RF beat signal is  amplified with an automatic-gain-control (AGC) amplifier and sent to a frequency counter (K+K Messtechnik, FXM50). A fraction of the RF-signal is used to establish a phase-lock of the Ti:Sa laser to the frequency comb. To this end, the beat signal is amplified again and fed to a phase-frequency detector (PFD) (Analog Devices, HMC403) where $\nu_{b}$ is compared to a \SI{60}{\mega\hertz} local oscillator. The error signal is transmitted to the control box of the ring laser \cite{haubrich96a} via an isolation amplifier (IA). The frequency components in the range $0\--$\SI{20}{\mega\hertz} are isolated with a diplexer, pre-amplified and distributed to an inverting bipolar high-voltage amplifier (APEX Microtechnology, PA90) and an amplifier (Comlinear, CLC103). The amplified signals are applied to the intracavity EOM as shown in Fig. \ref{fig:Ti:Sa_lock}. This frequency-offset-locking scheme provides a phase lock of the Ti:Sa ring laser to the ultra-low-noise frequency comb and makes $\nu_{\text{L}}$ SI traceable.\\
\subsection{Detection of the \textit{nkm}$\leftarrow$ 2s transition}
The $nkm\leftarrow 2{\rm s}$ excitation is carried out in the center of two electro-polished stainless-steel plates separated by $\approx \SI{2.1}{\centi\meter}$ and designed for the application of homogeneous electric fields. A ring electrode consisting of four segments is inserted between the two plates to eliminate all line-of-sight trajectories of charged particles to insulators. This measure effectively prevents accumulation of charges near the excitation volume and is crucial to reduce stray electric fields. The segmented geometry enables one to apply transverse electric fields for stray-field compensation. A short plate distance of \SI{2.1}{\centi\meter} between the ion-repeller plate and the grounded extraction plate was chosen to be able to generate electric fields up to \SI{6}{\kilo\volt\per\centi\meter} in less than \SI{29}{\nano\second} with a home-built \SI{12.5}{\kilo\volt} low-noise high-voltage switch. With such fields, Rydberg states with principal quantum number as low as 20 can be efficiently field ionized (see Fig. \ref{fig:PFI_Stark_inversion} below).\par 
The electronic circuit was conceived to combine the high voltage pulse with low-noise DC potentials (2V$_{\text{DC}}$) on the repeller plate using a 20-bit digital-to-analogue low-noise voltage source. This enabled us to either minimize stray-electric-field components or to apply well-defined electric fields in the \textit{z} direction. The only openings in the electrode structure surrounding the photoexcitation region are 5-mm-diameter holes along the molecular-beam axis and 9-mm-diameter holes for the UV laser beam.\\
\subsection{Doppler-shift cancellation}\label{sec:AFR}
The inset of Fig. \ref{fig:Experimental_setup} schematically describes the photoexcitation geometry, where $\vec{v}$ is the H(2s)-atom velocity and $\vec{k}$ the wavevector of the forward-propagating (blue) and reflected (red) UV radiation. Any deviation $\delta\alpha$ from \SI{90}{\degree} of the angle between the laser beam and the supersonic beam leads to a first-order Doppler shift. To cancel this shift, we choose $\delta\alpha$ to be large enough so that the spectral lines from the forward-propagating and reflected UV laser beams do not overlap. In addition, a \SI{180}{\degree} reflection angle is enforced through an active-stabilization feedback system, based on a design introduced in Refs. \cite{beyer13a,beyer16c,wirthl21a}. This procedure resulted in a mirror-symmetric double-line profile with center at the first-order Doppler-free frequency \cite{beyer18a}. Choosing $\delta\alpha$ as close to zero as possible, as advocated in  Ref. \cite{beyer16c,wirthl21a}, turned out not to be practical in our case because the nonthermal nature of the H(2s)-atom velocity distribution made it challenging to extract the central frequency from the lineshapes under conditions where the fine structure is not fully resolved. \par
An aberration-free set of four antireflection-coated lenses \cite{lenssise} with an effective focal length of \SI{21.35}{\milli\meter} is used to collimate the diverging beam emerging from a pure-silica-core, polarization-maintaining, single-mode optical fiber (mode-field diameter \SI{2.3}{\micro\meter}), resulting in a parallel beam with a M$^2$ value of $\approx$ 1.02. The focus of the resulting Gaussian beam is located $\approx$ \SI{20}{\meter} beyond the chamber. Consequently, the reflected beam almost exactly retraces the incoming beam and the change of wavefront curvature is negligible. \par
The active stabilization of the alignment of the \SI{180}{\degree} reflecting mirror is achieved by dithering its tip and tilt angles by applying sinusoidal electric potentials to piezo-electric elements installed at the back of the mirror holder (see Fig. \ref{fig:Experimental_setup}). The dithering leads to a modulation of the incoupling efficiency of the reflected beam into the silica-core fiber beyond the lens system. These modulations are detected with an auto-balanced photodiode (PD). The dithering frequencies are selected to minimize cross talk between the motions of the tip and tilt axes. The error signal used to correct the mirror position is produced by lock-in amplifiers (LIA) (Femto, LIA-MVD-200L) connected to a proportional-integral controller (PI). To compensate slow drifts, the time constant of the feedback loop was chosen to be \SI{0.1}{\second}.\par
\section{Theoretical description of Rydberg states of the H atom in electric fields}\label{sec:Stark_theory}
\begin{figure}[h!]
	\includegraphics[scale=1.0]{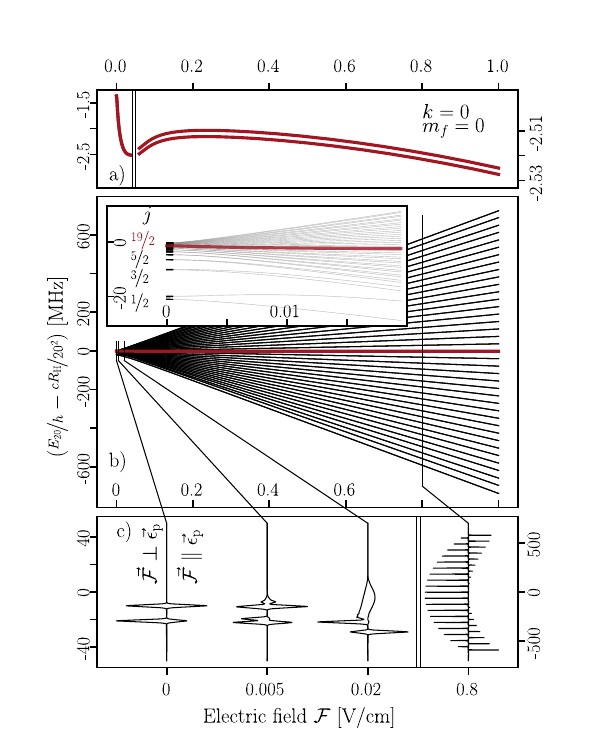}
	\caption{Stark effect in the $n=20,\,m_f=0$ manifold of the H atom. a) Field dependence of the $k=0$ state revealing a quadratic shift below \SI{50}{\milli\volt\per\centi\meter} caused by the intramanifold mixing of different orbital-angular-momentum components, and a smaller quadratic shift at larger fields arising from the interaction between different $n$ manifolds. b) Overview of the field dependence of all $m_l=0$ Stark states, which is essentially linear. c) Calculated spectra for different electric-fields strengths and electric-field vectors $\vec{\mathcal{F}}$ pointing parallel or perpendicular to the laser polarization $\vec{\epsilon}_{\mathrm{p}}$. }
	\label{fig:Stark_eff_m0}
\end{figure}

The energy levels of the H atom in a static homogeneous electric field $\vec{\mathcal{F}} = (0,0,\mathcal{F})$ are eigenvalues of the Hamiltonian
\begin{equation}\label{Eq:H_extField}
	\hat{\mathcal{H}}= \hat{\mathcal{H}}_0  + e\mathcal{F}\hat{z}, % \hat{\mathcal{H}}_0 + \hat{\mathcal{H}}' =
\end{equation}
where $\hat{\mathcal{H}}_0$ is a diagonal matrix containing the field-free energies of the $\left|nljfm_f\right>$ states with principal quantum number $n$, orbital angular momentum quantum number $l$, total angular momentum quantum number without nuclear spin $j$, total angular momentum quantum number $f$, and associated magnetic quantum number $m_f$. The field-free hyperfine-centroid energies, including terms arising from relativistic, quantum-electrodynamics (QED) and finite-nuclear-size corrections, can be accurately calculated using Eqs. $7\--41$ of Ref. \cite{tiesinga21a} and the latest recommended physical constants (2018 CODATA, see Ref. \cite{tiesinga02a}). To obtain the field-free energy-level structure at high $n$ values, we used Bethe logarithms tabulated in Ref. \cite{jentschura05a} and included the hyperfine splittings using the analytical expressions provided in Ref. \cite{horbatsch16a}. The calculated structure of the $m_l=0$ levels at $n=20$ is depicted in the inset of Fig. \ref{fig:Stark_eff_m0}b).\par
The operator $e\mathcal{F}\hat{z}$ in Eq. \ref{Eq:H_extField} describes the effect of the external field. The perturbation can be treated in excellent approximation in a nonrelativistic framework and relativistic corrections to the Stark effect as discussed in Ref. \cite{cordle74a} become negligible as $n$ increases. $e\mathcal{F}\hat{z}$ only contributes off-diagonal elements connecting zero-field states differing in $l$ by $\pm1$. These matrix elements can be expressed in analytic form using standard angular-momentum algebra (see, e.g., Refs. \cite{zare88a,sobelman92a}) as
\begin{align}\nonumber
	&\left<  n'l'j'f'm_f'\right| \hat{z} \left| nljfm_f\right> =(-1)^{\Delta f+\Delta j+\Delta l-m_f'+I+S}\times\\ \nonumber
	&\ThreeJ{l'}{1}{l}{0}{0}{0}\ThreeJ{f'}{1}{f}{-m_f'}{0}{m_f}\SixJ{j'}{f'}{I}{f}{j}{1}\SixJ{l'}{j'}{S}{j}{l}{1} \times\\
	&\sqrt{\Theta(f')\Theta(f)\Theta(j')\Theta(j)\Theta(l')\Theta(l)} \braket{n'l'}{r}{nl},
\end{align}
where the expressions in parentheses and curly parentheses are Wigner 3j and 6j symbols, respectively, $\Theta(x) = 2x+1$, $\Delta x = x'-x$, and $\braket{n'l'}{r}{nl}$ are radial integrals connecting the $r$-dependent parts of the solutions of the Schr\"odinger equation of the H atom (see Eqs. 63.2 and 63.5 of Ref. \cite{bethe57a}). Restricting the calculations of the Stark effect to a single $n$ value, one obtains an intra-manifold quadratic Stark effect at low fields and a linear Stark effect at intermediate fields, as depicted in Fig. \ref{fig:Stark_eff_m0}. The Stark states are commonly labeled by the parabolic quantum numbers $n_1$ and $n_2$ or by their difference $k = n_1 - n_2$ \cite{bethe57a,gallagher94a}. At intermediate field strengths, the states can approximately be described by their $k$ and $m_l$ values. States of a given value of $k$ form near degenerate groups with $m_l$ values ranging from $-(n - |k| - 1)$ to $(n-|k| -1)$ in steps of 2. The $k=0$ states, highlighted in red in Fig. \ref{fig:Stark_eff_m0}, are the only states retaining almost pure parity $\left[(-1)^{n-1}\right]$. They have a zero electric dipole moment and are insensitive to the field over a large range of fields, which makes them attractive for precision measurements, except at fields very close to zero. All other states exhibit a dipole moment in the field. At intermediate to high field strengths, the coupling between states of different $n$ values induced by the field becomes significant and the states start exhibiting an inter-manifold quadratic Stark effect. This behavior is displayed on an enlarged vertical scale for $m_f=0$ in Fig. \ref{fig:Stark_eff_m0}a). To reliably calculate Stark shifts in this field range, it is necessary to include basis states of neighboring $n$ values until convergence with the size of the basis set is reached.\par
\begin{figure}[h!]
	\includegraphics[scale=1.1]{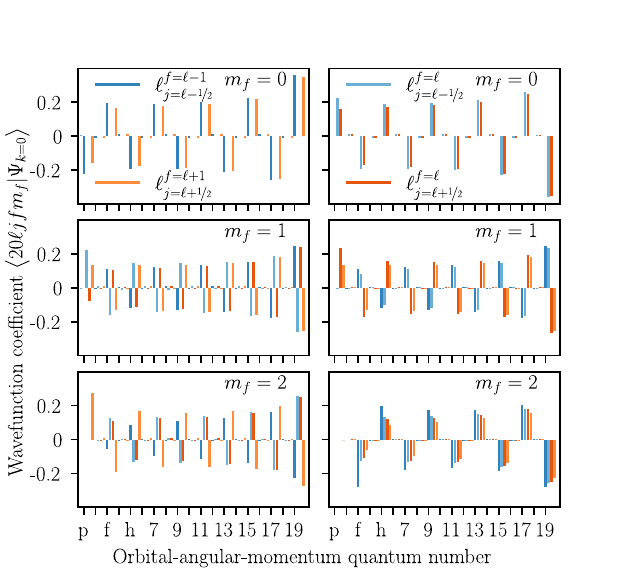}
	\caption{Expansion coefficients of the $k=0$, $|m_f|=0,1$ and 2 Rydberg-Stark wavefunctions in the $\left|ljfm_f\right>$ angular-momentum basis as labeled in the figure. Only basis states with odd orbital angular momentum quantum number make significant contributions.}
	\label{fig:l_character}
\end{figure}
Figure \ref{fig:l_character} presents the decomposition of the $ n=20,\,k=0$ Stark states with $m_f=0-2$ in the $\left|ljfm_f\right>$ basis. For each $m_f$ value, the eigenstates possess contributions from up to four hyperfine-structure components, as indicated by the color labels. The intensity of transitions from the 2s level corresponds to the coherent squared sum of the p characters in the evaluation of electric-dipole-moment matrix elements.\par
Figure \ref{fig:Stark_eff_m0}c depicts calculated intensity distributions in spectra of the $n=20 \leftarrow 2s$ transitions at field strength below \SI{1}{\volt\per\centi\meter} and for laser polarizations parallel and perpendicular to the DC electric field. At fields below \SI{20}{\milli\volt\per\centi\meter}, corresponding to typical stray fields, the center of gravity of the distribution depends on the polarization and varies strongly and nonlinearly with the field strength, making precision measurements prone to systematic uncertainties. This behavior explains why high-$n$ Rydberg states are usually avoided in precision measurements. However, in the linear regime of the Stark effect, \textit{i.e.}, above \SI{0.2}{\volt\per\centi\meter} at $n=20$, the spectra regain a regular intensity pattern and the spacings between the Stark states encode the field strength. When the polarization is parallel to the field ($\pi$ transitions), the intensity is strongest at the outer edges of the manifold and vanishes at $k=0$, for even $n$ values because $k=0, m_l=0$ states do not exist, and for odd $n$ values because $k=0$ states have vanishing p character. When the polarization is perpendicular to the field, the opposite behavior is observed (see right panel of Fig. \ref{fig:Stark_eff_m0}c).\par
Consideration of Fig. \ref{fig:Stark_eff_m0} leads to the following conclusions concerning precision spectroscopy in high-$n$ states of  hydrogen-like systems:
\begin{itemize}
	\item Because of the nontrivial field dependence of the line profiles, precision measurements are not attractive in the region of the intra-manifold quadratic Stark effect.
	\item In the linear regime of the Stark effect, regular spectral pattern are restored and the states with $|k|\ge0$ form pairs of levels with Stark shifts of opposite sign. The positions of the $k\neq0$ states can be used for the electric-field calibration, as will be demonstrated in Section \ref{sec:Result}.
	\item If an easily calculable shift from the Bohr energy $\left(-hcR_{\mathrm{H}}n^{-2}\right)$ arising from the quadratic Stark effect is disregarded, the $k=0$ Stark states are essentially field-independent. Consequently, spectra of $k=0$ Stark states in the linear regime are not subject to broadening by inhomogeneous fields and their positions can be converted into the Bohr energy by adding the calculated Stark shift (see red curves in Fig. \ref{fig:Stark_eff_m0}a)).
	\item The linear Stark manifold is thus perfectly suited for metrological purposes, in particular for precise determination of the Bohr energy. It has previously been used to determine the binding energy of Rydberg states of H$_2$ \cite{hoelsch22a}.
\end{itemize}
\begin{figure}[h!]
	\includegraphics[scale=1.0]{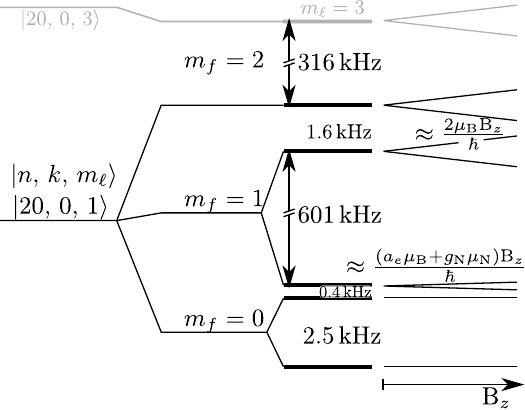}
	\caption{Energy level structure of the eight $n=20,\,k=0$ Stark states with $m_l=1$ character, calculated at an electric field strength $\mathcal{F}= $ \SI{0.8}{\volt\per\centi\meter}.  These states split into two groups of four states each separated by $\approx \SI{600}{\kilo\hertz}$. The Zeeman effect induced by a magnetic field pointing along the quantization axis is  schematically illustrated on the right side and lifts all remaining degeneracies.}
	\label{fig:Stark_HFS_Zeeman}
\end{figure}
The wavefunctions of the Stark states can be used to estimate their magnetic moments and systematic shifts arising from the Zeeman effect caused by residual magnetic fields, as illustrated in Fig. \ref{fig:Stark_HFS_Zeeman} with the example of the $k=0, |m_l|=1$ Stark states. In this case, the electric field splits the structure into two $m_f=0$, two $m_f=1$ and one $m_f=2$ components and a total of eight states. The magnetic moments are given by the relative orientations of the electron orbital angular momentum, electron spin, and nuclear spin vectors. A magnetic field parallel to the electric field further splits these components according to their magnetic moments, as displayed schematically on the right-hand side of Fig. \ref{fig:Stark_HFS_Zeeman}. Because the Zeeman shifts are symmetric and extremely small in a magnetically shielded environment (less than \SI{2.4}{\kilo\hertz} for $\mu = 2\mu_{\text{B}}$ and $|$B$|\leq$\SI{100}{\nano\tesla}), we conclude that the Zeeman effect in low-$m_l$ states can be neglected in metrological applications relying on Stark states in the linear regime. This is also the case for perpendicular magnetic-field components because the corresponding Zeeman effect couples states with $\Delta m_l =\pm1$ which are located in different $k$ manifolds and thus energetically too distant for significant mixing to occur.\par
As explained in Section \ref{sec:experimental_setup}, the maximal electric-field strength we apply to record Stark spectra is \SI{2}{\volt\per\centi\meter}. The applied fields also induce shifts of the 2s level energies, which need to be considered when extracting the absolute positions of the Rydberg-Stark states. The Stark shifts of the 2s levels can be calculated in the same manner as explained above for higher \textit{n} values.
\begin{figure}[h!]
	\includegraphics[scale=1.0]{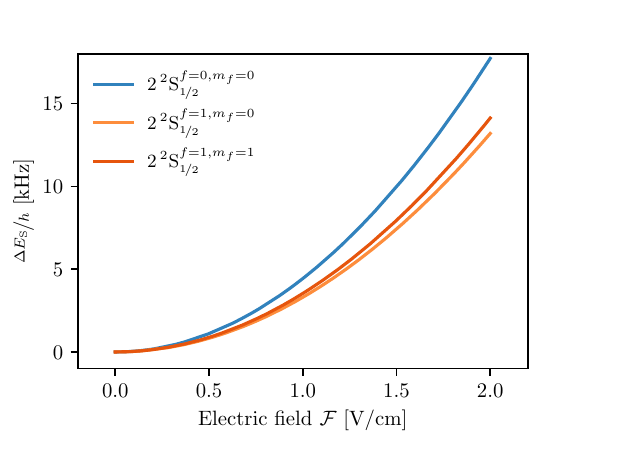}
	\caption{Stark shifts of the metastable 2s levels of the H atom calculated for electric fields in the range between 0 and \SI{2}{\volt\per\centi\meter}.}
	\label{fig:2s_Stark_shifts}
\end{figure}
The calculated shifts are displayed in Fig. \ref{fig:2s_Stark_shifts}. They are positive and quadratic for small electric fields because the dominant interactions are with the 2p$_{\nicefrac{1}{2}}$ states, which are located energetically just below the 2s states. When determining the absolute positions of the $nkm$ Rydberg-Stark states from spectra of the $nkm \leftarrow$ 2s transitions, the 2s Stark shifts must be added to the measured transition frequencies.
\section{Results}\label{sec:Result}
\begin{figure}[h!]
	\includegraphics[scale=1.0]{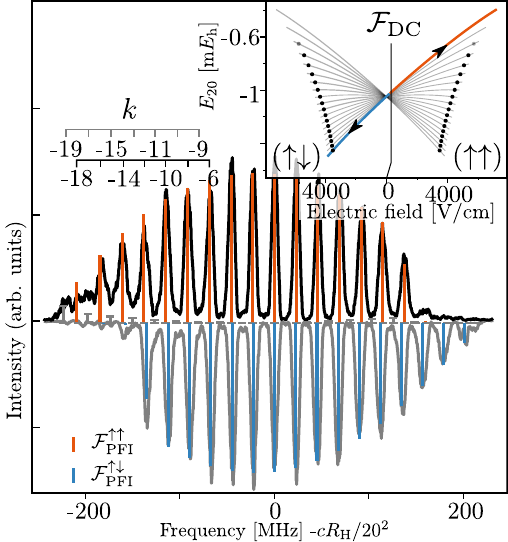}
	\caption{PFI spectra of the $n=20$ Rydberg-Stark states of H recorded from the 2s($f=1$) hyperfine component in  an electric field $\mathcal{F}_{\mathrm{DC}}\approx\SI{200}{\milli\volt\per\centi\meter}$. The direction of the strong pulsed electric-field ($\mathcal{F}_{\text{PFI}}=\SI{5.7}{\kilo\volt\per\centi\meter}$) used for ionization was set parallel to $\mathcal{F}_{\mathrm{DC}}^{\uparrow\uparrow}$ to record the upper spectrum and antiparallel $\mathcal{F}_{\mathrm{DC}}^{\uparrow\downarrow}$ to record the lower, inverted spectrum. The red and blue stick spectra represent the calculated intensity distributions.  Inset: The alignment of the two fields leads to ionization without change of the field polarity (red) or to ionization after a diabatic state inversion upon reversal of the field polarity. }
	\label{fig:PFI_Stark_inversion}
\end{figure}
Figure \ref{fig:PFI_Stark_inversion} displays pulse-field-ionization (PFI) spectra of the $n=20$ Stark manifold recorded from the 2s($f=1$) hyperfine level using laser radiation polarized linearly in the direction orthogonal to the applied DC electric field $\mathcal{F}_{\text{DC}}$. The upper (lower) trace was recorded by field ionizing the Rydberg states with a pulsed field $\mathcal{F}_{\text{PFI}}$ pointing in the same (opposite) direction as the DC field $\left[\mathcal{F}_{\text{PFI}}= \SI{5.7}{\kilo\volt\per\centi\meter}, \mathcal{F}_{\text{DC}} = \SI{0.2}{\volt\per\centi\meter} (\SI{-0.2}{\volt\per\centi\meter} )\right]$. The orthogonal laser-polarization arrangement led to the observation of dominant transitions to Stark states of even $k$ values, as assigned at the top of the figure. The intensity distributions in both spectra are very similar, except at the edges of the manifold. Whereas the intensities of the transitions to the highest $k$ states ($k\geq14$) are strongly depleted in the upper spectrum, the lowest $k$ states ($k\leq -14$) are depleted in the lower spectrum. The reason for the disappearance of the intensities at the edges of the Stark manifold are twofold: First, the transition dipole moment gradually decreases with increasing $|k|$ value. Second, the ionization rates of the Stark states that are shifted to higher energies by the pulsed field rapidly decrease with increasing $k$ value. In the case of the upper spectrum, these states are those observed at the highest frequencies. For the lower spectrum, they are observed at the lowest frequencies because of the reversal of the sign of $k$ when the field polarity changes upon application of the pulsed field, which diabatically inverts the Stark manifold, as schematically illustrated in the inset. This interpretation is fully supported by calculations of the spectral intensities, as depicted in the red and blue stick spectra in Fig. \ref{fig:PFI_Stark_inversion}. These intensities were obtained by multiplying the squared transition dipole moments calculated as explained in Section \ref{sec:Stark_theory} with the field-ionization probabilities over the 80-ns-long detection window calculated using the analytical expressions reported by Damburg and Kolosov \cite{damburg79a}. \par
\begin{figure}[h!]
	\includegraphics[scale=1.0]{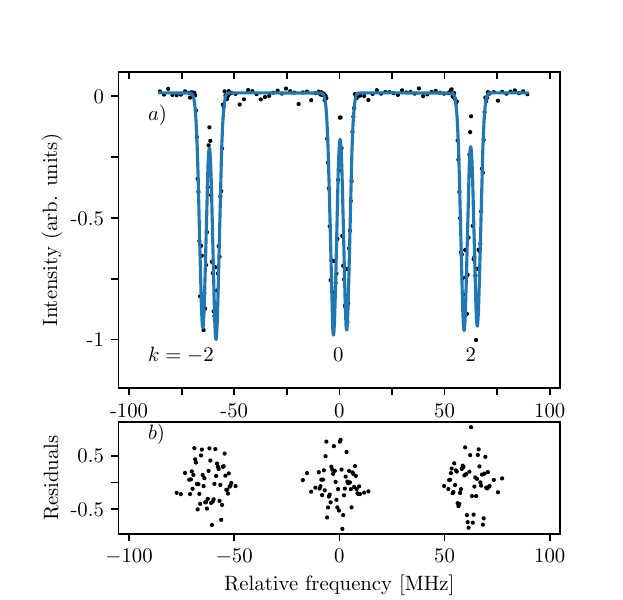}
	\caption{a) Typical experimental (dots) and fitted (blue) spectra of the three ($k = 0,\pm2$) Rydberg-Stark states near the center of the $n=20$ Stark manifold of H,  each exhibiting two Doppler components. b) Weighted residuals (see text for details). }
	\label{fig:Stark_spectrum}
\end{figure}
Before recording the lower spectrum in Fig. \ref{fig:PFI_Stark_inversion}, the transverse stray fields were carefully compensated. Consequently, the laser polarization was almost perfectly parallel to the DC field. Under these conditions, transitions to Stark states of odd $k$ values have zero intensity. In the case of the upper spectrum, a weak transverse stray field made the Stark states with odd $k$ values optically accessible. Transitions to these states are strongest at the edges of the manifold and weakest at the center. The calculated intensities of transitions to odd $k$ states in the presence of the transverse stray field ($\sim \SI{10}{\milli\volt\per\centi\meter}$) are depicted as gray sticks in Fig. \ref{fig:PFI_Stark_inversion}. They are only observable at the low-frequency edge of the Stark manifold because the Stark states at the high-frequency edge are not efficiently ionized by the pulsed field, as explained above. The good agreement between measured and calculated intensity distributions enables us to conclude that the Rydberg-Stark states located near the center of the $n=20$ manifold are fully ionized by the \SI{5.7}{\kilo\volt\per\centi\meter} pulsed field used in the experiments.\par
Figure \ref{fig:Stark_spectrum} displays a typical spectrum of transitions to the $k=0,\pm2$ Stark states of the $n=20$ manifold recorded from the H(2s,$f=1$) state using laser radiation with linear polarization orthogonal to the \SI{0.8}{\volt\per\centi\meter} DC field. The spectrum was recorded at an angle deviation $\delta\alpha=\SI{1.1}{\milli\radian}$ from exact orthogonality between the H-atom beam and the laser beam, leading to two Doppler components per $k$ state, separated by \SI{6.28}{\mega\hertz}. The two Doppler components are slightly asymmetric with mirror-symmetric lineshapes (opposite sign of $\gamma$ in Eq.~\ref{Eq:Lineshape} below). To optimize the data acquisition rate when recording the Stark spectra, the frequency was scanned in steps of \SI{400}{\kilo\hertz} within the line profiles and of \SI{2}{\mega\hertz} between the lines. In addition, the data points within the spectral lines were obtained by averaging over 500 experimental cycles (\textit{i.e.}, over \SI{20}{\second}) whereas only 100 cycles were averaged for data points between the lines. The central frequency, the electric field strength and additional parameters were determined in a least-squares fit to the experimental data (black dots) based on the following line profile for each $k$ value
\begin{widetext}
	\begin{align}\label{Eq:Lineshape}
		&g_k(\nu) = \sum_{i=1}^{2}\sum_{m_f = -2}^{2} \mathrm{I}^i \mathrm{I}^{m_f}(\mathcal{F})\exp\left\{\frac{-\left[\nu -\nu_0^{i,m_f}(\mathcal{F},\gamma)\right]^2}{2\left(\sigma_{\mathrm{D}}^2 +|k| \sigma_{\mathrm{S}}^2\right)}\right\}\times\left[1 + \mathrm{erf}\left((-1)^i\gamma\frac{\left(\nu -\nu_0^{i,m_f}(\mathcal{F},\gamma)\right)}{\sqrt{2}\sigma_{\mathrm{D}}}\right)\right],
	\end{align}
\end{widetext}
with
\begin{align}\label{Eq:Lineshape_2}
	&\nu_0^{i,m_f}(\mathcal{F},\gamma) =  \nu_0 + \nu_{\mathrm{S}}^{m_f}(\mathcal{F}) + (-1)^i\left\{\nu_{\mathrm{D}} -\delta\nu(\gamma)\right\}.
\end{align}
In Eqs. \ref{Eq:Lineshape} and \ref{Eq:Lineshape_2}, $i\, (=1,2)$ is an index specifying the Doppler component, $\nu_0$ is the transition frequency to the reference position $\left(-\nicefrac{cR_{\text{H}}}{n^2}\right)$ of the calculated Stark map of the $n=20$ levels (see Fig. \ref{fig:Stark_eff_m0}), $\nu_{\mathrm{S}}^{m_f}(\mathcal{F})$ is the field-dependent Stark shift of the $m_f$ level, $\nu_{\mathrm{D}}$ is the Doppler shift arising from the angle deviation $\delta\alpha$, and $\delta\nu(\gamma)$ is a frequency offset used to compensate the shift of the intensity maximum of the asymmetric line profiles from the centers of the hypothetical symmetric profiles. This shift is introduced to reduce the correlation between the asymmetry parameter $\gamma$ and $\nu_{\mathrm{D}}$ in the least-squares fit. $\sigma_{\mathrm{D}}$ is the Doppler width and $\sigma_{\mathrm{S}}$ accounts for the broadening of the $|k|=2$ lines arising from weak field inhomogeneities in the photoexcitation volume. As mentioned in Section \ref{sec:2s_excitation}, the asymmetry of the line profiles originate from the nonthermal velocity distribution caused by the 2s $\leftarrow$ 1s excitation.\\
\begin{table}[h]
	\caption{Fit results obtained in the least-squares fit of the line profiles based on Equations~\ref{Eq:Lineshape} and~\ref{Eq:Lineshape_2}.}\label{tab:Fitresults_1}\centering
	\begin{tabular}{cc}
		\hline
		\hline
		&value\\
		\hline
		$\nicefrac{\nu_0}{\text{kHz}}$	& \num{0\pm26} (blinded)\\
		$\nicefrac{\mathcal{F}}{\text{V\,cm$^{-1}$}}$&  \num{0.8076\pm0.0003}\\
		$\nicefrac{\nu_{\mathrm{D}}}{\text{MHz}}$& \num{3.16\pm0.05} \\
		$\nicefrac{\sigma_{\mathrm{D}}}{\text{MHz}}$& \num{1.56\pm0.07}\\
		$\nicefrac{\sigma_{\mathrm{S}}}{\text{MHz}}$& \num{0.27\pm0.06} \\
		$\gamma$& \num{0.65\pm0.18} \\
		\hline
	\end{tabular}
\end{table}
The fit of the line profiles depicted in Fig. \ref{fig:Stark_spectrum} resulted in the parameters listed in Table \ref{tab:Fitresults_1}. These parameters are helpful in characterizing the experimental conditions. For instance, the homogeneous component of the field is found to correspond closely to the \SI{0.8}{\volt\per\centi\meter} applied experimentally with an uncertainty of only $0.4\permil$ or \SI{300}{\micro\volt\per\centi\meter}. The electric field inhomogeneity leads to a broadening of the $k=\pm2$ Stark components and is well represented by a field gradient of \SI{12\pm3}{\milli\volt\per\centi\meter\squared}, which corresponds to a field change of \SI{2.4\pm0.6}{\milli\volt\per\centi\meter} over the \SI{2}{\milli\meter} diameter of the UV laser. The Doppler shift $\nu_{\mathrm{D}}$ reflects the deviation angle $\delta\alpha$ which, in this case, is \SI{1.1}{\milli\radian}. $\sigma_{\mathrm{D}}$ is a measure of the transversal velocity distribution, which in the present case corresponds to a temperature of \SI{40}{\micro\kelvin} and is the result of the geometric constraints along the supersonic beam imposed by the skimmers and the $2\mathrm{s}\leftarrow1\mathrm{s}$ excitation. The asymmetry parameter is alignment specific and typically varied between -2 and 4. The central frequency was arbitrary set to zero because the absolute frequency determination is still in a blinded phase. The weights used for the least-squares fits are determined in an iterative procedure to approach a normal distribution of the residuals. \par
\begin{figure}[h!]
	\includegraphics[scale=1.0]{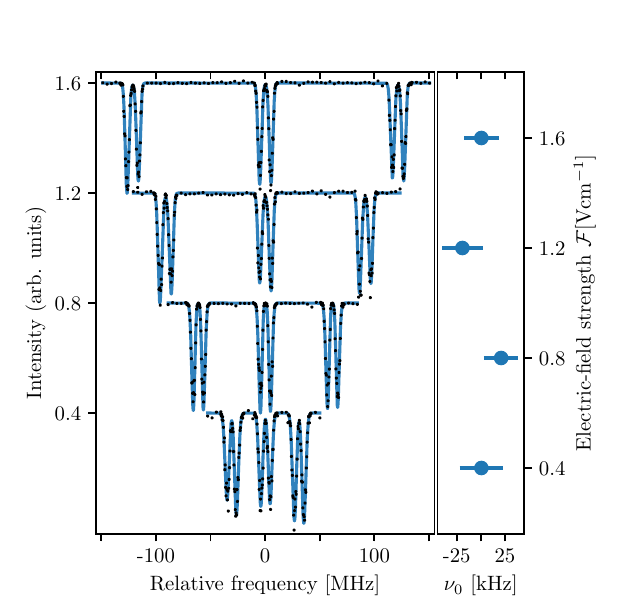}
	\caption{Spectra of the $n=20, k = 0,\,\pm2,\,|m_l|=1 \leftarrow 2\text{s}(f=1)$ transitions measured when applying nominal electric fields of 0.4, 0.8, 1.2 and \SI{1.6}{\volt\per\centi\meter}, respectively. Each spectrum represents the sum of three independent scans as described in Section \ref{sec:experimental_setup}. Right: Relative positions of the line center $\nu_0$ with respect to the line center measured at a nominal field strength of \SI{0.4}{\volt\per\centi\meter}. The error bars represent 1$\sigma$ uncertainties.}
	\label{fig:Stark_Field_spectra}
\end{figure}
The overall data set collected so far involves more than 500 individual spectra of transitions recorded from the initial 2s$(f=1)$ and 113 from the 2s$(f=0)$ hyperfine state to $n=20$ Rydberg states and 35 spectra from the 2s$(f=1)$ to $n=24$ Rydberg states. These spectra were recorded for different valve temperatures, electric-field strengths and deviation angles $\delta\alpha$ to investigate possible sources systematic uncertainties.\\
The main objective of the study presented here was to verify that the central frequencies extracted from the spectra do not depend on the strength of the applied electric field. A typical set of four measurements recorded at nominal field strengths of 0.4, 0.8, 1.2 and \SI{1.6}{\volt\per\centi\meter}  under otherwise identical experimental conditions (beam velocity of \SI{1060}{\meter\per\second} and deviation angle $\delta\alpha$ of \SI{1.1}{\milli\radian}) is presented in Fig. \ref{fig:Stark_Field_spectra}. At the scale of the figure, the Stark effect appears essentially linear. Table \ref{tab:Stark_Field_spectra} summarizes the relevant lineshape parameters (see Eqs. \ref{Eq:Lineshape} and \ref{Eq:Lineshape_2}) extracted from the fits of the lineshapes to the experimental data.\par
\begin{table}[h]\center
	\caption{Lineshape parameters extracted from fits to the measured spectra of the $n=20, k = 0,\,\pm2,\,|m_l|=1 \leftarrow 2\text{s}(f=1)$ transitions measured when applying a nominal electric field of 0.4, 0.8, 1.2 and \SI{1.6}{\volt\per\centi\meter}, respectively. }\label{tab:Stark_Field_spectra}
	\begin{tabular}{ c c c c c}
		\hline
		\hline
		&\SI{0.4}{\volt\per\centi\meter} & \SI{0.8}{\volt\per\centi\meter} & \SI{1.2}{\volt\per\centi\meter}& \SI{1.6}{\volt\per\centi\meter}\\
		\hline
		$\nicefrac{\nu_0}{\mathrm{kHz}}$&0(21)&21(18)&-20(21)&-0(20)\\
		$\nicefrac{\mathcal{F}}{\mathrm{Vcm}^{-1}}$&0.4012(4)&0.7990(3)&1.1882(3)&1.5794(3)\\
		$\nicefrac{\sigma_{\mathrm{S}}}{\mathrm{MHz}}$&0.31(10)&0.22(9)&0.17(14)&0.23(6)\\
		$\nicefrac{\nu_{\mathrm{D}}}{\mathrm{MHz}}$&4.26(5)&4.61(4)&5.20(5)&5.18(3)\\
		$\nicefrac{\sigma_{\mathrm{D}}}{\mathrm{MHz}}$&2.02(10)&1.78(5)&2.12(5)&1.81(5)\\
		\hline
	\end{tabular}
\end{table}
The central frequencies corrected for the Stark shift of the 2s state agree within the combined uncertainties and do not reveal any systematic dependence on the field strength within the \SI{20}{\kilo\hertz} accuracy of the measurements. The field strength corresponds to the applied electric potential within the expected uncertainties resulting from the geometry of the electrode plates and the electronic circuits used to apply the potentials. The field-dependent line broadening does not reveal a significant dependence on the applied field strength, which suggests that the applied field distribution does not contribute to the observed field inhomogeneity. The slight variations in the values of $\nu_{\mathrm{D}}$ and $\sigma_{\mathrm{D}}$ reflect small changes in the day-to-day alignments of the beams and the supersonic-beam properties.\par

\begin{figure}[h!]
	\includegraphics[scale=1.0]{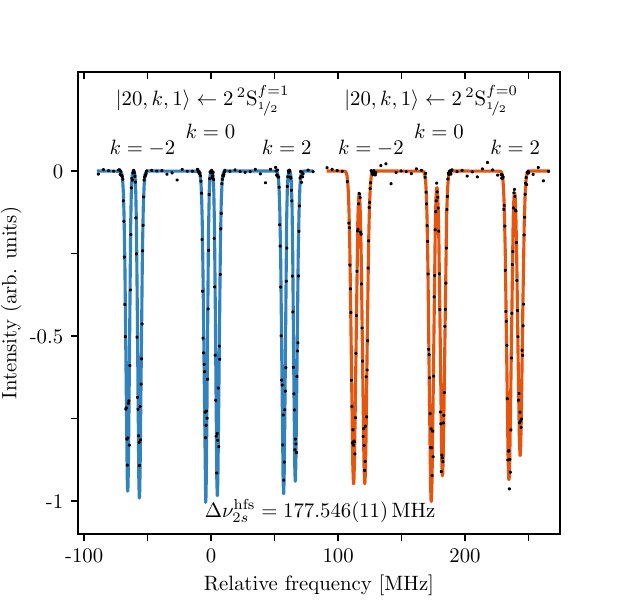}
	\caption{Spectra of the $n=20, k = 0,\,\pm2,\,|m_l|=1 \leftarrow 2\text{s}(f=1)$ (blue)  and $n=20, k = 0,\,\pm2,\,|m_l|=1 \leftarrow 2\text{s}(f=0)$ (red). The difference of the two central frequencies $\nu_0$ corresponds to the hyperfine interval of the 2s state.}
	\label{fig:f1f2_interval}
\end{figure}
\begin{figure}[h!]
	\includegraphics[scale=1.0]{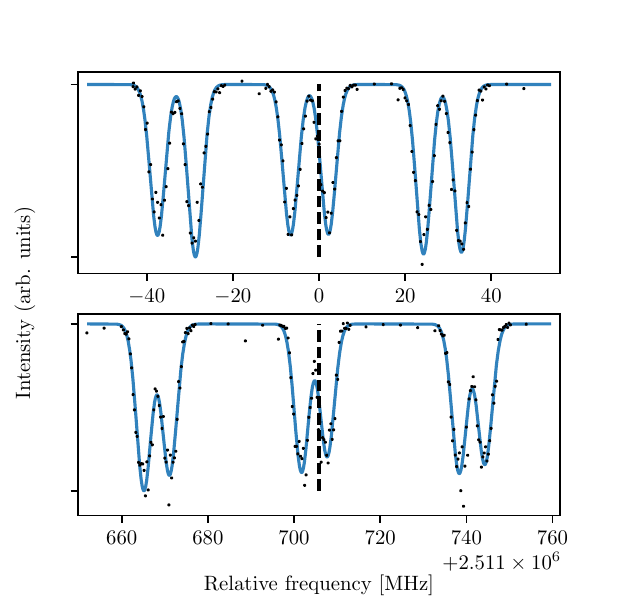}
	\caption{Spectra of the $n=20, k = 0,\,\pm2,\,|m_l|=1 \leftarrow 2\text{s}(f=1)$ (top)  and $n=24, k = 0,\,\pm2,\,|m_l|=1 \leftarrow 2\text{s}(f=1)$ (bottom). The energy scale has its origin at the $n=20$ Bohr energy which is located $\approx \SI{2.2}{\mega\hertz}$ above the $n=20,k=0$ Stark states (see Fig. \ref{fig:Stark_eff_m0}). The Bohr energies are indicated by the dashed lines in the two panels.}
	\label{fig:n20n24_interval}
\end{figure}

The data set collected so far was used to determine the hyperfine splitting in the 2s level as well as the difference between the Bohr energies of the $n=20$ and $n=24$ Rydberg states. Figure \ref{fig:f1f2_interval} presents spectra of the transitions to the $n=20$, $k = 0,\pm2$ Stark states recorded from the 2s$(f=0)$ (red) and 2s$(f=1)$ (blue) states as illustration. Taking the difference in the central frequencies $\nu_0$ (see Eq. \ref{Eq:Lineshape_2}) for the two sets of data (197 spectra and 50 spectra for $f=1$ and $f=0$, respectively) yields a value of \SI{177.546\pm0.011}{\mega\hertz} for the 2s hyperfine splitting, which corresponds within the 1$\sigma$ uncertainty to the much more precise value of 177.55683887(85)\,MHz determined by Ramsey spectroscopy in the $n=2$ manifold \cite{bullis23a}.\\
The difference in the Bohr energies of the $n=20$ and 24 Rydberg states was determined in an analogous manner from spectra of the $n=20$ and 24 Stark states recorded from the 2s$(f=1)$ state as illustrated in Fig. \ref{fig:n20n24_interval}.
The difference of the two $\nu_0$ values is \SI{2511705.793\pm0.010}{\mega\hertz} and also agrees within the experimental uncertainty with the value $cR_{\mathrm{H}}\left(\nicefrac{1}{20^2} - \nicefrac{1}{24^2}\right) =\SI{2511705.802}{\mega\hertz}$. The uncertainty of \SI{10}{\kilo\hertz} results from the addition in quadrature of the \SI{7}{\kilo\hertz} uncertainties of the blinded $\nu_0$ values extracted from the experimental data.
\section{Conclusion}
In this article, we have outlined an experimental approach to determine $R_{\infty}$ from $k=0,\,\pm2,\,|m_l|=1$ Rydberg-Stark spectra of H. We have demonstrated that systematic errors resulting from the Stark effect are insignificant within the $\sim \SI{11}{\kilo\hertz}$ precision of the four data sets used as illustrations (see Fig. \ref{fig:Stark_Field_spectra}). We have also demonstrated that the differences between the Bohr energy at $n=20$ and the positions of the $f=0$ and 1 hyperfine components of the 2s state are consistent within the \SI{11}{\kilo\hertz} statistical uncertainty of the present determination with the more precise value of the 2s $(f=0) \--( f=1)$ interval determined recently by Ramsey microwave spectroscopy \cite{bullis23a}.\\
Finally, we have determined the difference between the Bohr energies at $n=20$ and 24 and found the results to agree with Bohr's formula using the CODATA 2018 recommended value for $R_{\mathrm{H}}$ \cite{tiesinga21a}. The data presented in this article was collected over a period of several months with frequent realignment of the optical system and supersonic beam. We did not observe inconsistencies in any of the relative frequencies determined for this article over this time.\\
The 2s$(f=0)\--(f=1)$ and $\nu_0(n=24)\--\nu_0(20)$ intervals presented in this article correspond to differences of large frequencies, and systematic errors largely cancel out when building the differences. The main potential source of systematic errors in our method originates from the Doppler effect and a possible imperfect cancellation of the Doppler shifts. To characterize such uncertainties, measurements of absolute frequencies are underway, in which we systematically vary the velocity of the supersonic beam and the deviation angle $\delta\alpha$. Absolute transition frequencies will be reported when these measurements are completed.\\
%

\begin{acknowledgments}
We thank Dominik Husmann (METAS, Bern) for his help in maintaining the SI-traceable frequency dissemination network and Gloria Clausen for helpful discussions. We also thank Prof. Klaus Ensslin and Peter M\"arki for the low-noise DC voltage source used in the measurements of the Stark spectra.\\
This work was supported by the Swiss National Science Foundation through the Sinergia-program (Grant No. CRSII5-183579) and a single-investigator grant (Grant No. 200020B-200478).
\end{acknowledgments}

%\bibliography{mybib}
%\bibliography{mybib,grpbib}

\begin{thebibliography}{60}
\bibitem [{\citenamefont {\relax Lens~assembly}()}]{lenssise}%
	\BibitemOpen
	\bibfield  {author} {\bibinfo {author} {\bibnamefont {\relax
				Lens~assembly}},\ }\href@noop {} {}\bibinfo {note} {\relax $d$:distance of
		the first surface to the previous lens; $R_1$ and $R_2$: radius of the two
		surfaces; and $CT$: center thickness. L1=$\left[ d=7.96, R_1=25.50,CT=5.06,
		R_2=13.27\right]$, L2=$\left[ d=2.1, R_1=25.49, CT=5.1, R_2=13.28\right]$,
		L3=$\left[d=11.3,R_1=-36.46, CT=6.11, R_2=30.60\right]$, L4=$\left[ d=4.2,
		R_1=17.21, CT=4.12, R_2=30.59\right]$. All dimensions are in mm}\BibitemShut
	{NoStop}%
\end{thebibliography}

\end{document}